\documentclass[letterpaper]{article} % DO NOT CHANGE THIS
\usepackage{arXiv}  % DO NOT CHANGE THIS
\usepackage{times}  % DO NOT CHANGE THIS
\usepackage{helvet}  % DO NOT CHANGE THIS
\usepackage{courier}  % DO NOT CHANGE THIS
\usepackage[hyphens]{url}  % DO NOT CHANGE THIS
\usepackage{graphicx} % DO NOT CHANGE THIS
\usepackage{natbib}  % DO NOT CHANGE THIS AND DO NOT ADD ANY OPTIONS TO IT
\usepackage{caption} % DO NOT CHANGE THIS AND DO NOT ADD ANY OPTIONS TO IT
\usepackage{algorithm}
\usepackage{algorithmic}

\usepackage{xcolor}
\usepackage{makecell}
\usepackage{multicol}
\usepackage{booktabs}
\usepackage{multirow}
\usepackage{xspace}
\usepackage{amsmath}
\usepackage{textcomp}

\usepackage{verbatimbox}

\newtheorem{definition}{Definition}

\newcommand{\LLMs}[0]{\textsc{LLMs}\xspace}
\newcommand{\LLM}[0]{\textsc{LLM}\xspace}
\newcommand{\LLMCs}[0]{\textsc{LLMCs}\xspace}
\newcommand{\LLMC}[0]{\textsc{LLMC}\xspace}

\newcommand\AST[0]{\textsc{AST}\xspace}

\newcommand{\Granite}[0]{\textsc{Gra\-ni\-te\-Co\-de}\xspace}
\newcommand{\SemCoder}[0]{\textsc{Sem\-Co\-der}\xspace}

\newcommand{\Mistral}[0]{\textsc{Mis\-tral}\xspace}

\newcommand{\Qwen}[0]{\textsc{Qwen\-2.5-Co\-der}\xspace}
\newcommand{\NewQwen}[0]{\textsc{Qwen\-3-Co\-der}\xspace}

\newcommand{\CodeGemma}[0]{\textsc{Co\-de\-Gem\-ma}\xspace}
\newcommand{\Llama}[0]{\textsc{Lla\-ma\-3.2}\xspace}
\newcommand{\CodeLlama}[0]{\textsc{Co\-de\-Lla\-ma}\xspace}

\newcommand{\LiveCodeBench}[0]{\textsc{LiveCodeBench}\xspace}
\newcommand{\CruxEval}[0]{\textsc{CruxEval}\xspace}

\newcommand{\Gemini}[0]{\textsc{Gemini\--3}\xspace}
\newcommand{\GPT}[0]{\textsc{GPT\--5.2}\xspace}

\usepackage[newfloat,frozencache,cachedir=.]{minted}
\usepackage{tikz}
\usetikzlibrary{arrows,automata,backgrounds,calc,chains,external,fit,graphs,patterns,positioning,shapes,shapes.geometric}

\tikzset{
    >=stealth',
    punkt/.style={
           rectangle,
           rounded corners,
           draw=black, very thick,
           text width=10em,
           minimum height=2em,
           text centered},
    pil/.style={
           ->,
           thick,
           shorten <=2pt,
           shorten >=2pt,},
    ionode/.style={
           rectangle,
           rounded corners,
           text width=12em,
           minimum height=4em,
           text centered},
}

\tikzset{fancy/.style={rectangle,
		rounded corners=1mm,
		ultra thin,
		draw=white,
		top color=white,
		bottom color=black!20,
		draw}}

\definecolor{highClr}{rgb}{1.0, 1.0, 0.0}

\colorlet{edgeClr}{orange!80!black}
\tikzset{sandEdge/.style={
		>=stealth,
		shorten >=1pt,
		thick,
		bend left,
		text=black,
		edgeClr,
	}}

\tikzset{fadedEdge/.style={
		->,
		>=stealth,
		shorten >=1pt,
		thick,
		edgeClr!20,
	}}

\tikzset{weigthLabel/.style={
		text=black,
		sloped,
		midway,
		}}

\tikzset{fadedWeigth/.style={
		text=lightgray!40,
		sloped,
		midway,
		anchor=south,
		}}

\tikzset{blueVertex/.style={
		rectangle,minimum size=6mm,rounded corners=3mm,
		top color=white,bottom color=blue!35!cyan!25!,
		font=\ttfamily,
		text=black,
	}}

\tikzset{blueVertexG/.style={
		rectangle,minimum size=6mm,rounded corners=3mm,
		top color=white,bottom color=blue!35!cyan!25!,
		font=\ttfamily,
		text=black,
		draw=green,
		thick,
	},
	blueVertexY/.style={
		rectangle,minimum size=6mm,rounded corners=3mm,
		top color=white,bottom color=blue!35!cyan!25!,
		font=\ttfamily,
		text=black,
		draw=yellow,
		thick,
	},
	blueVertexO/.style={
		rectangle,minimum size=6mm,rounded corners=3mm,
		top color=white,bottom color=blue!35!cyan!25!,
		font=\ttfamily,
		text=black,
		draw=orange,
		thick,
	}}

\colorlet{noteClr}{lightgray!30!white!50}

\tikzset{noteBckg/.style={
		rounded corners=8pt,fill=noteClr,
	},
	noteStl/.style={
		font=\scriptsize,
		align=center,
		text=black
	}}

\makeatletter

\newcommand*\mysize{%
  \@setfontsize\mysize{8}{8}%
}
\makeatother
\setminted{fontsize=\mysize}

\title{Are Large Language Models Robust in Understanding Code Against Semantics-Preserving Mutations?}

\author{
    Pedro Orvalho,
    Marta Kwiatkowska
}
\affiliations{
    Department of Computer Science\\University of Oxford\\Oxford, UK
}

\begin{document}

\maketitle

\begin{abstract}
With the widespread adoption of \emph{vibe coding}, understanding the reasoning and robustness of Large Language Models (\LLMs) is critical for their reliable use in programming tasks. While recent studies assess \LLMs’ ability to predict program outputs, most focus on accuracy alone, without evaluating the underlying reasoning.
Moreover, it has been observed on mathematical reasoning tasks that \LLMs can arrive at correct answers through flawed logic, raising concerns about similar issues in code understanding. 
In this paper we assess whether state-of-the-art \LLMs  can reason about Python programs or are simply guessing.
We apply five semantics-preserving code mutations: renaming variables, mirroring comparison expressions, swapping if-else branches, converting \texttt{for} loops to \texttt{while}, and loop unrolling.
These mutations maintain program semantics while altering its syntax.
We evaluated nine \LLMs, including both open-source and closed-access models, and performed a human expert analysis using \LiveCodeBench to assess whether correct predictions are based on sound reasoning. We also evaluated prediction stability across different code mutations on \LiveCodeBench and \CruxEval. 
While proprietary models achieve the strongest predictive accuracy and reasoning quality in the expert evaluation, our robustness analysis reveals substantial fragility under semantics-preserving transformations.
Our findings show that \LLMs trained for code produce correct predictions based on flawed reasoning in between 10\% and 50\% of cases. Furthermore, \LLMs often change predictions in response to our code mutations, with performance drops reaching up to 70\%, indicating that they do not yet exhibit stable, semantically grounded reasoning, even when initial accuracy is high.
\end{abstract} 

\section{Introduction}

Large Language Models~(\LLMs) have rapidly become integral to a wide range of daily tasks, from writing assistance to code generation. In particular, the software development community has embraced \LLM-based tools, such as \textsc{GitHub Copilot} and \textsc{ChatGPT}, to streamline code workflows, assist in debugging, and even automate code completion and review~\cite{neurips24-SemCoder,icse24-Liang0M24}. These tools are widely used, and often blindly, with developers placing significant trust in their capabilities~\cite{sp24-llms-trust}.
However, this growing reliance on \LLMs for coding tasks raises a fundamental question: To what extent do \LLMs truly understand code and the underlying semantics of programs? 

While recent \LLMs can produce syntactically correct code and even solve competitive programming problems, there is a risk that their responses may reflect pattern recognition over code syntax rather than genuine semantic understanding~\cite{proof-or-bluff-LLMs-on-Math}. 
Although \LLMs rely on statistical associations to generate output, their ability to mimic reasoning can obscure the limitations of such models, potentially leading to an overestimation of their reliability in critical development contexts~\cite{acl24-conterfeit-conundrum}.
Furthermore, based on the growing dependence of programmers on \LLMs, several works in the past year have analysed \LLMs' reasoning about code through output prediction tasks~\cite{neurips24-SemCoder,gu2024cruxeval,livecodebench}. Some \LLMs, such as \SemCoder~\cite{neurips24-SemCoder}, were trained to \emph{understand} program semantics through a monologue reasoning strategy~(i.e., Chain-of-Thought), 
where the model verbally simulates both the high-level syntax and low-level execution effects of code. However, some recent work has shown that in other domains, such as mathematical competitions~\cite{proof-or-bluff-LLMs-on-Math}, \LLMs tend to provide accurate predictions, 
but based on flawed~reasoning.

In this paper, we evaluate \LLMs on the task of reasoning about code and predicting program outputs for given inputs, using two well-studied benchmarks: \LiveCodeBench~\cite{livecodebench} and \CruxEval~\cite{gu2024cruxeval}. We define \emph{sound semantic reasoning} in \LLMs in terms of their reasoning traces being semantically accurate, that is, when the models predict the correct output and correctly describe the control and data flow that takes the program from the inputs to the expected output. 
First, we conduct a manual expert evaluation on \LiveCodeBench to determine whether correct predictions are derived from logically sound reasoning, flawed reasoning, or mere guesses. Second, we assess model robustness by applying semantics-preserving code mutations to both benchmarks and analysing output prediction consistency. These mutations, which are small syntactic modifications that preserve runtime behaviour, test whether \LLMs’ understanding of code is robust to syntactic changes.
We apply five semantics-preserving code mutations explained in Section~\ref{sec:code-mutations}: (1) renaming variables, (2) mirroring comparison expressions, (3) swapping if-else branches, (4) converting \texttt{for} loops to \texttt{while} loops, and (5) unrolling the final iterations of loops. 
These code transformations allow us to evaluate whether \LLMs are sensitive to syntax or capable of reasoning about~code~semantics.

%%%%%%%%%%%%%%%%%%%%%%%%%%%%%%%%%%%%%%%%%%%%%%%%%%%%%%%%%%%%%%%%%%%%%%%%
\begin{table*}[t!]
\begin{minipage}[t!]{0.47\linewidth}
\centering
\begin{algorithm}[H]
\caption{Function \texttt{minPossibleSum(n, target)}.}
\label{code:min-possible-sum}
\begin{minted}[escapeinside=||,tabsize=0,obeytabs,xleftmargin=0pt,fontsize=\footnotesize]{python}
def minPossibleSum(n:int, target:int)->int:
    i = 1
    arr = {1}
    while len(arr) < n:
        i += 1
        if target - i not in arr:
            arr.add(i)
    return sum(arr)
\end{minted}
\end{algorithm}
\end{minipage}
% \hfill
\hspace{0.2in}
\begin{minipage}[t!]{0.47\linewidth}
\centering
\begin{algorithm}[H]
\caption{Algorithm~\ref{code:min-possible-sum} after renaming variable \texttt{arr}.}
\label{code:min-possible-sum-renamed}
\begin{minted}[escapeinside=||,tabsize=0,obeytabs,xleftmargin=0pt,fontsize=\footnotesize]{python}
def minPossibleSum(n:int, target:int)->int:
    i = 1
    eAJMfVcq = {1}
    while len(eAJMfVcq) < n:
        i += 1
        if target - i not in eAJMfVcq:
            eAJMfVcq.add(i)
    return sum(eAJMfVcq)
\end{minted}
\end{algorithm}
\end{minipage}
\end{table*}
%%%%%%%%%%%%%%%%%%%%%%%%%%%%%%%%%%%%%%%%%%%%%%%%%%%%%%%%%%%%%%%%%%%%%%%%

Our experiments consider both open-source and closed-access state-of-the-art \LLMs.
We evaluate nine different \LLMs: six models specifically trained for coding tasks, \CodeGemma, \Granite, \Mistral, \Qwen, \NewQwen, and \SemCoder, two closed-access models that are widely used as coding agents, \Gemini and \GPT, and one general-purpose model, \Llama, included for comparison.
Our in-depth analysis of \LLMs' reasoning about code shows that some \LLMs produce correct answers based on flawed reasoning in up to 55\% of cases for general-purpose \LLMs, and 50\% for \LLMs trained specifically for coding tasks~(\LLMCs). 
Additionally, we observe that interactive querying enhances both prediction accuracy and reasoning quality, increasing the rate of sound reasoning by 14\% in \LLMCs and 19\% in \Llama. Moreover, our experiments show that the evaluated \LLMs frequently alter their predictions under semantics-preserving code mutations, by up to 70\%, indicating that these models do not yet exhibit stable, semantically~grounded~reasoning. 
Variable renaming consistently causes substantial changes in model predictions across both datasets, despite preserving program semantics, indicating that identifier names strongly influence model behaviour, a sensitivity also observed in human comprehension~\cite{ese19-var-renaming}. In contrast, loop unrolling leads to the largest performance drops, highlighting limited structural and semantic generalisation across models.

Our expert evaluation shows that the strongest performance was achieved by \GPT and \Gemini, which attained near-perfect prediction accuracy and consistently sound semantic reasoning. \NewQwen also demonstrated particularly strong reasoning quality among the open-access models. Nevertheless, when evaluated under semantics-preserving code mutations, even these top-performing models exhibit substantial fragility. In fact, \GPT and \Gemini experience some of the largest performance drops across mutations, revealing that high accuracy on the original benchmark does not necessarily imply semantic robustness.
These results must therefore be interpreted in light of \emph{model release timelines} and accessibility constraints. The evaluated versions of \GPT and \Gemini correspond to releases from the end of 2025, whereas \LiveCodeBench was published in 2024. In contrast, most open-access models in our study were released prior to the benchmark’s publication. Consequently, \NewQwen, \GPT, and \Gemini may have been trained on \LiveCodeBench or overlapping data, potentially inflating their performance. 
Although the exact composition of proprietary training corpora is unknown, the near-perfect results of \GPT and \Gemini, and the strong performance of \NewQwen, should therefore be interpreted with caution.
Performance differences must also be considered alongside \emph{accessibility}. While the open-access models evaluated in this work are freely available, \GPT and \Gemini require paid API access. Thus, although proprietary models achieve substantially higher predictive accuracy and reasoning quality in standard evaluations, this advantage comes with a non-trivial cost differential and does not translate into robustness against semantics-preserving transformations.

Thus, our experiments serve as evidence that semantics-preserving perturbations expose the brittleness of purely data-driven reasoning, and merely scaling models or training on more code will not fix this problem because the flaw is rooted in the \emph{absence of formal semantic constraints}.
Reliable reasoning about code demands \emph{formal guarantees that current \LLMs cannot provide}. While current state-of-the-art code-focused models excel at pattern recognition and surface-level syntax manipulation, our findings show that they fail to maintain stable predictions under semantics-preserving transformations, revealing a fundamental lack of true semantic understanding. 

In summary, this paper makes the following contributions:
\begin{itemize}
\item We perform a manual expert evaluation to assess whether \LLMs' code output predictions are based on logically sound code reasoning, flawed reasoning, or mere~guesses.
\item We evaluate \LLMs' output prediction stability across five  different semantics-preserving code mutations.
\item Experiments show that \LLMs frequently alter their predictions when subjected to semantics-preserving code mutations, highlighting that they do not yet exhibit stable, semantically grounded~reasoning.
\item We discuss fruitful research directions that integrate formal methods with machine learning to create semantics-aware code models, highlighting how such hybrid approaches could improve robustness and security against adversarial code~mutations.
% \item Our code and datasets will be made publicly available on \textsc{GitHub}~(see~supplementary~material).
% ~\cite{figshare-anon-code} .
\end{itemize}

\section{Motivation}
\label{sec:motivation}

Consider the function presented in Algorithm~\ref{code:min-possible-sum}. 
Given two positive integers \texttt{n} and \texttt{target}, function \texttt{minPossibleSum} finds the sum of an array \texttt{arr} that meets the following conditions:
\texttt{len(arr) == n}, all elements in \texttt{arr} are distinct positive integers, and there does not exist two distinct indices, \texttt{i} and \texttt{j}, in the range \texttt{[0, n - 1]}, such that \texttt{arr[i] + arr[j] == target}. This corresponds to the programming contest \#2834 from \textsc{LeetCode}~\cite{leetcode}.

Now imagine we ask a Large Language Model (\LLM) trained for coding tasks (\LLMC), such as \Qwen~\cite{Qwen-2024}, to predict the output of the function \texttt{minPossibleSum(n=1, target=1)} and to explain its reasoning using the prompt described in Section~\ref{sec:interactive-process}. \Qwen produces flawed reasoning and an incorrect output prediction, replying as follows: 

\begin{verbnobox}[\fontsize{9pt}{9pt}\selectfont]
Let's walk through the execution of the code 
step-by-step to understand how the output 
is produced.
\end{verbnobox}

\begin{verbnobox}[\fontsize{9pt}{9pt}\selectfont]
The function `minPossibleSum` takes 
two parameters:  `n` and `target`. 
In this case, `n = 1` and `target = 1`.
\end{verbnobox}

\begin{verbnobox}[\fontsize{9pt}{9pt}\selectfont]
1. The variable `i` is initialized to 1.
2. A set `arr` is initialized with the 
element 1.
3. The while loop condition checks if 
the length of `arr` 
is less than `n`. Since `n` is 1, the 
loop will run once.
4. Inside the loop, `i` is incremented 
to 2.
5. The condition `if target - i not in 
arr` is checked. 
Here, `target - i` is `1 - 2`, which 
equals `-1`. 
Since `-1` is not in the set `arr`, the 
condition is true.
6. The number 2 is added to the set `arr`.
7. The loop ends because the length of 
`arr` is now `n`.
8. The function returns the sum of the 
elements in `arr`, which is `1 + 2 = 3`.
\end{verbnobox}

\begin{verbnobox}[\fontsize{9pt}{9pt}\selectfont]
Therefore, the assertion should be:
[ANSWER]assert minPossibleSum(1, 1) == 3[/ANSWER]
\end{verbnobox}

The predicted output is incorrect, as the semantically correct output is \texttt{1}. Moreover, the reasoning provided by the model regarding the function’s semantics is flawed. When we feed this feedback back into \Qwen using a feedback prompt described in Section~\ref{sec:interactive-process}, it %\Qwen 
generates the following revised response:

\begin{verbnobox}[\fontsize{9pt}{9pt}\selectfont]
There seems to be a misunderstanding in the 
explanation. Let's re-evaluate the logic:
- Initially, `arr` contains `{1}`.
- The loop runs while the length of `arr` 
is less than `n`.
- In each iteration, `i` is incremented, 
and if `target - i` is not in `arr`, `i` 
is added to `arr`.
\end{verbnobox}

\begin{verbnobox}[\fontsize{9pt}{9pt}\selectfont]
Given `n = 1` and `target = 1`:
- Initially, `arr = {1}`.
- The loop does not run because `len(arr) < n` 
(1 < 1) is false.
- Therefore, the loop does not add any 
new elements to `arr`. Thus, the final 
`arr` remains `{1}`, and the sum is `1`.
Therefore, the correct assertion should be:
[ANSWER]assert minPossibleSum(1, 1) == 1[/ANSWER]
\end{verbnobox}

Thus, our feedback was sufficient to guide \Qwen toward a logically sound reasoning process and to correctly predict the output. In contrast, \SemCoder, another \LLM trained for coding tasks, was able to produce the correct result along with logically sound reasoning on the first attempt.
However, when these two \LLMs are presented with the semantically equivalent function shown in Algorithm~\ref{code:min-possible-sum-renamed}, where the variable \texttt{arr} has been renamed to \texttt{eAJMfVcq}, both models fail to provide the correct output and fail to construct logically sound reasoning, even after five rounds of interactive feedback.
This experiment highlights a key limitation in the semantic robustness of \LLMs trained for code: their reasoning processes are fragile and often disrupted by small syntactic changes, even when program semantics is preserved. Additionally, this experiment suggests that interactive querying can be an effective strategy to guide \LLMs toward correcting flawed~reasoning. 

Please refer to Appendix~\ref{sec:interaction-llms} for additional examples of interactions with the evaluated~\LLMs.

%%%%%%%%%%%%%%%%%%%%%%%%%%%%%%%%%%%%%%%%%%%%%%%%%%%%%%%%%%%%%%%%%%%%%%%%

\section{Preliminaries}

This section provides definitions used throughout the paper.

\begin{definition}[Abstract Syntax Tree (AST)]
\label{def:AST}
An~\emph{abstract syntax tree (AST)} is a syntax tree in which each node represents an operation, and the children of the node represent the arguments of the operation for a given language described by a context-free grammar~\cite{hopcroft2008introduction}.
An \AST represents the grammatical structure of a program~\cite{compilers-book-dragon}. 
\end{definition}

\begin{definition}[Semantics-Preserving Code Mutation.]
Given $(T, G, O, P)$, let $T$ be a set of input-output examples (test suite), $G$ be a grammar, $O$ be the semantics for a particular domain-specific language (DSL), and $P$ be a syntactically well-formed program (i.e., a set of statements, instructions, expressions) consistent with $G$ and $O$, such that $P$ is semantically consistent with the test suite, i.e., \newline
% \[
$\forall{(t^i_{in},t^i_{out}) \in T}\ :\ P(t^i_{in}) = t^i_{out}$.
% \]

A \emph{semantics-preserving code mutation} is a syntactic program transformation to $P$ that generates a new program $P_m$ by syntactically replacing a subset $S_1$ of $P$’s statements ($S_1 \subseteq P$) with another set of statements $S_2$ consistent with $G$ and $O$, such that
$P_m = ((P \setminus S_1) \cup S_2)$,
and $P_m$ is semantically consistent with the original specification: \newline
% \[
$\forall{(t^i_{in},t^i_{out}) \in T}\ :\ P_m(t^i_{in}) = t^i_{out}$.
% \]
\end{definition}

In Section~\ref{sec:code-mutations}, we introduce a set of five semantics-preserving program transformations designed to syntactically modify Python programs without altering their semantics, i.e., their runtime behaviour: variable renaming, comparison~mirroring, swap if-else statements, for-to-while loop conversion, and partial loop unrolling. These code mutations are essential for tasks such as testing the robustness of code understanding models and augmenting training data in a semantics-preserving manner.

\begin{table*}[t!]
\hfill
\begin{minipage}[t!]{0.32\linewidth}
\centering
% \vspace{0.15in}
\begin{algorithm}[H]
\caption{Original Python program.}
\label{code:original}
% \KwIn{None}
% \KwOut{Maximum of three numbers}
% ,linenos
\begin{minted}[escapeinside=||,tabsize=0,obeytabs,xleftmargin=0pt]{python}
def f(nums):
  sum = 0
  for n in nums:
    if n % 2 == 0:
      sum += n  
    else:
      sum += 0
  return sum
\end{minted}
\end{algorithm}
\end{minipage}
\hfill
\hfill
\begin{minipage}[t!]{0.32\linewidth}
\centering
% \vspace{0.15in}
\begin{algorithm}[H]
\caption{Renaming of variable \texttt{sum}.}
\label{code:rename_vars}
% \KwIn{None}
% \KwOut{Maximum of three numbers}
\begin{minted}[escapeinside=||,tabsize=0,obeytabs,xleftmargin=0pt]{python}
def f(nums):
  uoWIfiQc = 0
  for n in nums:
    if n % 2 == 0:
      uoWIfiQc += n  
    else:
      uoWIfiQc += 0
  return uoWIfiQc
\end{minted}
\end{algorithm}
\end{minipage}
\hfill
\hfill
\begin{minipage}[t!]{0.32\linewidth}
\centering
% \vspace{0.15in}
\begin{algorithm}[H]
\caption{Mirroring if-condition.}
\label{code:mirror-expr}
% \KwIn{None}
% \KwOut{Maximum of three numbers}
\begin{minted}[escapeinside=||,tabsize=0,obeytabs,xleftmargin=0pt]{python}
def f(nums):
  sum = 0
  for n in nums:
    if 0 == n % 2:
      sum += n  
    else:
      sum += 0
  return sum
\end{minted}
\end{algorithm}
\end{minipage}
\hfill
\hfill
\begin{minipage}[t!]{0.32\linewidth}
\centering
% \vspace{0.15in}
\begin{algorithm}[H]
\caption{Swapping if-else statements.}
\label{code:swap-if-else}
% \KwIn{None}
% \KwOut{Maximum of three numbers}
\begin{minted}[escapeinside=||,tabsize=0,obeytabs,xleftmargin=0pt]{python}
def f(nums):
  sum = 0
  for n in nums:
    if not n % 2 == 0:
      sum += 0  
    else:
      sum += n
  return sum
\end{minted}
\end{algorithm}
\end{minipage}
\hfill
% \hfill
\begin{minipage}[t!]{0.32\linewidth}
\centering
% \vspace{0.45in}
\begin{algorithm}[H]
\caption{Converting for-to-while loop.}
\label{code:for-2-while}
% \KwIn{None}
% \KwOut{Maximum of three numbers}
\begin{minted}[escapeinside=||,tabsize=0,obeytabs,xleftmargin=0pt]{python}
def f(nums):
  sum = 0
  i = 0
  while i < len(nums):
    n = nums[i]
    if n % 2 == 0:
      sum += n  
    else:
      sum += 0
    i += 1
  return sum
\end{minted}
\end{algorithm}
\end{minipage}
% \hfill
\hfill
\begin{minipage}[t!]{0.32\linewidth}
\centering
\vspace{-0.25cm}
\begin{algorithm}[H]
\caption{Partial loop unrolling.}
\label{code:unroll-loop}
% \KwIn{None}
% \KwOut{Maximum of three numbers}
\begin{minted}[escapeinside=||,tabsize=0,obeytabs,xleftmargin=0pt]{python}
def f(nums):
  sum = 0
  i = 0
  while i < (len(nums)-1):
    n = nums[i]
    if n % 2 == 0:
      sum += n  
    else:
      sum += 0
    i += 1
  if len(nums) > i:
    n = nums[i]
    if n % 2 == 0:
      sum += n  
    else:
      sum += 0
    i += 1
  return sum
\end{minted}
\end{algorithm}
\end{minipage}
\hfill
\end{table*}

%%%%%%%%%%%%%%%%%%%%%%%%%%%%%%%%%%%%%%%%%%%%%%%%%%%%%%%%%%%%%%%%%%%%%%%%

\section{Semantics-Preserving Code Mutations}
\label{sec:code-mutations}

We introduce a set of semantics-preserving program transformations designed to syntactically modify Python programs without altering their semantics, i.e., their runtime behaviour. These code mutations are essential for tasks such as testing the robustness of code understanding models and augmenting training data in a semantics-preserving manner. In this section, we describe five mutations implemented in our work: variable renaming, comparison~mirroring, swap if-else statements, for-to-while loop conversion, and partial loop unrolling. 
Some of these code mutations have been previously employed~\cite{saner22-robustness-models-code,ase19-refactory,fse22-multIPAs} to augment benchmarks with semantically equivalent, yet syntactically different, versions of the original programs.
As a running example, 
% throughout this section, 
we use the program shown in Algorithm~\ref{code:original}, where the function \texttt{f(nums)} returns the sum of all even numbers in the list~\texttt{nums}.

\textbf{Variable Renaming.}
This program transformation systematically renames local variables, function arguments, or function names using fresh identifiers that do not conflict with existing symbols or Python built-ins. A consistent mapping is maintained within each scope to ensure correctness. This program mutation preserves the program's semantics while altering its lexical structure.
Algorithm~\ref{code:rename_vars} shows the function \texttt{f(nums)} from Algorithm~\ref{code:original} after renaming the variable \texttt{sum} to \texttt{uoWIfiQc}. 
While variable renaming preserves syntax and semantics, it affects human comprehension~\cite{ese19-var-renaming}, making it a meaningful probe for~model~robustness.

\textbf{Comparison Expression Mirroring.}
This transformation mirrors comparison expressions by swapping operands and applying their logically equivalent inverse operators. 
This mutation preserves program semantics and applies to all symmetric and reversible binary comparisons. It is particularly useful for assessing models that rely on syntax, such as token order or abstract syntax tree (\AST) structure, for reasoning about code.
Algorithm~\ref{code:mirror-expr} shows the function \texttt{f(nums)} from Algorithm~\ref{code:original} after applying this mutation to change \texttt{n \% 2 == 0} into \texttt{0 == n \% 2}.

\textbf{Swap If-Else Statements.}
Another semantics-preserving mutation we use is the swapping of if-else statements, in which the \texttt{if} and \texttt{else} blocks are swapped and the \texttt{if} condition is logically negated. For example, a condition \texttt{if x > 0:} is rewritten as \texttt{if not (x > 0):}, with the bodies of the \texttt{if} and \texttt{else} blocks swapped accordingly. This code mutation maintains the program's behaviour but alters the logical structure and control flow graph~\cite{allen70-ControlFlowGraph}.
Robust models should recognise the semantic equivalence of these logically inverted blocks and produce consistent output predictions, regardless of the branching structure. Algorithm~\ref{code:swap-if-else} shows the function \texttt{f(nums)} from Algorithm~\ref{code:original} after applying this mutation. 

\textbf{For-to-While Loop Conversion.}
This code transformation rewrites \texttt{for} loops into semantically equivalent \texttt{while} loops. It introduces an index variable and manually iterates over the collection using the \texttt{len()} function and explicit indexing, while preserving all loop control logic. This transformation is particularly useful for assessing the robustness of \LLMs in recognising semantically equivalent loop constructs. Algorithm~\ref{code:for-2-while} shows the function \texttt{f(nums)} from Algorithm~\ref{code:original} after converting the \texttt{for} loop into a \texttt{while}~loop.

\textbf{Partial Loop Unrolling.}
To simulate partial loop unrolling while preserving semantics, we extract the last one or two iterations of a \texttt{while} loop body and duplicate them after the loop. The loop condition is also modified to run fewer iterations, e.g., reducing a bound \texttt{n} to \texttt{n - 1} in a \texttt{while i < n} condition. The extracted iterations are then executed sequentially after the loop, preserving the program behaviour. 
Algorithm~\ref{code:unroll-loop} shows the function \texttt{f(nums)} from Algorithm~\ref{code:for-2-while}, after converting the \texttt{for} loop into a \texttt{while} loop and then unrolling the last iteration.

\section{\LLM-Based Program Output Prediction}
\label{sec:interactive-process}
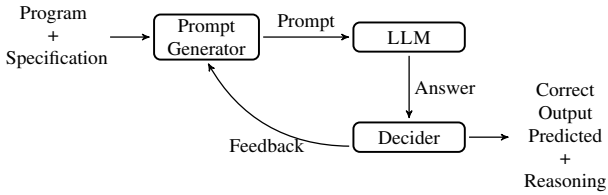
\begin{figure}[t!]
\centering
\resizebox{1\columnwidth}{!}{
\begin{tikzpicture}[node distance=2cm, auto,every node/.style={font=\fontsize{16}{16}\selectfont}]
 \node[punkt] (enum) {Prompt Generator};
 \node[punkt, inner sep=5pt,right=2cm of enum]
 (llm) {\LLM}
 edge[pil, <-] node[above]  (l) {Prompt} (enum.east);
 \node[punkt, inner sep=5pt,below=1.5cm of llm]
 (veri) {Decider} edge[pil,bend left=15] node[below=0.2cm] {Feedback} (enum.south)
 edge[pil, <-] node[right=0cm of llm]  (l) {Answer} (llm.south);
\node[punkt,draw=none,left=1cm of enum] (specs_node) {Program\\+\\Specification}
  edge[pil] (enum.west);
  \node[punkt,draw=none,right=1cm of veri] (output) {Correct Output Predicted\\+\\Reasoning}
  edge[pil, <-] (veri.east);
\end{tikzpicture}
}
\caption{\LLM-Based Program Output Prediction.}
\label{fig:output-prediction-loop}
\end{figure}

As illustrated in Figure~\ref{fig:output-prediction-loop}, we address the task of using \LLMs to predict the output of a Python program for a given input by employing an iterative querying strategy. 
We chose code output prediction as our evaluation task because success in this task strongly correlates with ``understanding'' of code semantics, but it is also easy to check and automate the interaction.
Starting with a program and its input-output specification, we invoke a prompt generator that constructs and submits the query to the \LLM. We then evaluate whether the predicted output matches the expected one. If the prediction is incorrect, a feedback prompt is generated and sent back to the \LLM, explicitly requesting a revised answer. This loop continues until the model produces the correct output, the time limit is reached, or the number of iterations exceeds five, typically indicating that the model is stuck and repeatedly returning the same incorrect answer.

%%%%%%%%%%%%%%%%%%%%%%%%%%%%%%%%%%%%%%%%%%%%%%%%%%%%%%%%%%%%%%%%%%%%%%%%

%%%%%%%%%%%%%%%%%%%%%%%%%%%%%%%%%%%%%%%%%%%%%%%%%%%%%%%%%%%%%%%%%%%%%%%%

\textbf{Prompts.} The prompts used to query the \LLMs follow a similar format to those adopted in prior works~\cite{neurips24-SemCoder,gu2024cruxeval}. Each prompt asks the model to complete a Python assertion, given the function signature and a test input. To guide the model's response format, we include a brief example demonstrating how the answer should be wrapped within specific tags,~as~follows:

% \begin{small}
% \begin{verbatim}
\begin{verbnobox}[\fontsize{9pt}{9pt}\selectfont]
Simulate the Execution: You are given a Python 
function and an assertion containing a function 
input. Complete the assertion containing the 
execution output corresponding to the given 
input in [ANSWER] and [/ANSWER] tags. 
For example, the answer to 
'''assert sumEvenNumbers([1,2,3,4])==???''' 
would be
[ANSWER]
assert sumEvenNumbers([1,2,3,4]) == 6
[/ANSWER]
Please complete the assertion and explain 
your reasoning for your prediction, using 
no more than 1000 tokens.
```python 
def f(nums):
    # python function
assert f([1, 2, 3, 4, 5]) == ???
```
\end{verbnobox}
% \end{verbatim}
% \end{small}

\textbf{Feedback.} If the output predicted by the \LLM is incorrect, i.e., it does not match the expected output from the test suite, we provide feedback to the model through iterative querying; this does not apply to two of the evaluated models (\SemCoder, and \Mistral) that do not support more than one interaction. Specifically, a follow-up prompt is sent indicating that the previous response was incorrect. This feedback aims to simulate a correction loop and assess whether the \LLM can refine its prediction after being notified of a mistake.
An example of such a feedback prompt~is:

\begin{verbnobox}[\fontsize{9pt}{9pt}\selectfont]
Your previous output prediction was INCORRECT!
Try again. Complete the initial program 
assertion containing the execution output 
corresponding to the given input in 
[ANSWER] and [/ANSWER] tags. 
For example, the answer to 
'''assert sumEvenNumbers([1,2,3,4])==???''' 
would be
[ANSWER]
assert sumEvenNumbers([1,2,3,4]) == 6
[/ANSWER]
Please complete the assertion and explain 
your reasoning for your new prediction, 
using no more than 1000 tokens.
\end{verbnobox}

%%%%%%%%%%%%%%%%%%%%%%%%%%%%%%%%%%%%%%%%%%%%%%%%%%%%%%%%%%%%%%%%%%%%%%%%

\section{Experiments}
\label{sec:results}

The goal of our experiments was to answer the following research questions (RQs):
% \\
\textbf{RQ1.} Are Large Language Models~(\LLMs) truly reasoning about code semantics, or merely guessing likely answers?
% \\
\textbf{RQ2.} Does the interactive querying process help \LLMs arrive at correct predictions supported by logically sound reasoning?
% \\
\textbf{RQ3.} Do different code mutations lead \LLMs to produce different predictions for the same program?
% \\
\textbf{RQ4.} Are \LLMs robust in understanding code against semantics-preserving mutations?
% \\
Our objective is to evaluate \LLMs on the task of reasoning about code and predicting program outputs for given inputs.
All experiments were conducted with a 5-minute time limit to bound each interaction loop described in Section~\ref{sec:interactive-process}.

\subsection{Experimental Setup}
All experiments were run using a NVIDIA L40S graphics card with 48GB of memory on an Intel(R) Xeon(R) Gold 5418Y 48-Core CPU Processor and 251GB RAM.

\textbf{Evaluation Benchmarks:}
To evaluate our approach, we use two widely adopted benchmarks of Python programs: \LiveCodeBench~\cite{livecodebench} and \CruxEval~\cite{gu2024cruxeval}.
\LiveCodeBench contains 479 programs submitted to programming contests across competition platforms, such as \texttt{LeetCode}~\cite{leetcode} and \texttt{CodeForces}~\cite{codeforces}. \CruxEval contains 800 functions generated by \CodeLlama~\cite{CodeLlama-LLM-2023}, each accompanied by a set of input-output examples for evaluation.
These benchmarks are commonly used to assess the capabilities of \LLMs across a range of code reasoning tasks, including test output prediction, program synthesis, and repair.
We apply each of the semantics-preserving code mutations described in Section~\ref{sec:code-mutations} to the programs in the benchmarks by randomly selecting applicable statements for transformation. We then verify that the semantics of the original program is preserved in the mutated version by checking that the program passes all assertions. For each code mutation, we generate up to two mutated variants per program, ensuring that each variant contains at most one~mutation.

\textbf{Large Language Models (\LLMs):} 
We evaluate nine different \LLMs: seven open-access \LLMs using the iterative querying setup described in Section~\ref{sec:interactive-process}, \CodeGemma~\cite{CodeGemma-LLM-2024}~(7B), \Granite~\cite{granite-LLM-2024}~(8B), \Mistral~\cite{Mistral-7B-2023}~(7B), \Qwen~\cite{Qwen-2024}~(7B), \NewQwen~\cite{qwen3technicalreport}~(30B), \SemCoder~\cite{neurips24-SemCoder}~(7B), and \Llama~\cite{Llama3-LLMs-2024}~(3B), and two closed-access models, OpenAI’s \GPT~\cite{chatgpt5.2} and Google’s \Gemini~\cite{gemini3}, which are among the most widely adopted state-of-the-art proprietary models.

\begin{table*}[t!]
\resizebox{2.1\columnwidth}{!}{%
\setlength{\tabcolsep}{1mm}
% \fontsize{9}{9}\selectfont
\centering
\begin{tabular}{lccccccccc}
\toprule
\textbf{Large Language Models} & \textbf{\CodeGemma} & \textbf{\Granite} & \textbf{\Qwen}  & \textbf{\NewQwen} & \textbf{\Mistral} & \textbf{\SemCoder} & \textbf{\Llama}  & \textbf{\GPT}  & \textbf{\Gemini} \\
\midrule
\textbf{\%Failed Predictions} & 66.8 & 65.1 & 37.4 & 49.5 & 67.4 & 52.0 & 61.4 & 2.5 & 0.0 \\ \midrule
\textbf{\%Correct Predictions} & 33.2 & 34.9 & 62.6 & 50.5 & 32.6 & 48.0 & 38.6 & 97.5 & 100.0 \\ \midrule \midrule
\begin{tabular}{@{}c@{}}\textbf{\%Correct Guesses}\\ \textbf{based on} \\\textbf{flawed reasoning}\end{tabular} & 44.7 & 43.1 & 10.3 & 3.0  & 50.0 & 16.1 & 55.1 & 0.2 & 0.2 \\ \midrule
\begin{tabular}{@{}c@{}}\textbf{\%Correct Predictions} \\ \textbf{based on} \\\textbf{sound reasoning}\\ \textbf{($>$1 iteration)}\end{tabular} & 3.8 & 13.8 & 8.0 & 1.6 & -- & -- & 18.9 & 8.6 & 0.0 \\ \midrule
\begin{tabular}{@{}c@{}}\textbf{\%Correct Predictions} \\ \textbf{based on} \\\textbf{sound reasoning}\\ \textbf{($=$1 iteration)}\end{tabular} & 51.6 & 43.1 & 81.7 & 95.4 & 50.0 & 83.9 & 25.9 & 91.2 & 99.8 \\
\bottomrule
\end{tabular}%
}
\caption{In-depth Analysis of \LLMs' Reasoning on \LiveCodeBench.}
\label{tab:prediction_breakdown}
\end{table*}

\subsection{Expert Analysis of \LLMs' Reasoning About Code}

To answer our first research question ({\bf RQ1}), Table~\ref{tab:prediction_breakdown} presents a detailed expert analysis of the reasoning behaviour of the evaluated \LLMs on \LiveCodeBench, focusing on prediction outcomes and code understanding. The lower section of Table~\ref{tab:prediction_breakdown} reports the proportion of correct predictions that were either guesses based on flawed reasoning or grounded in sound semantic reasoning, distinguishing between first-attempt success and success achieved after iterative refinement.

We define \emph{sound semantic reasoning} as reasoning traces that are semantically accurate, that is, when models not only predict the correct output but also correctly describe the control and data flow leading from the inputs to the expected output. All interactions between the evaluated \LLMs and programs in \LiveCodeBench were manually inspected. Labelling was conducted blind to model identity. Each chain-of-thought~(CoT) response leading to a correct prediction was independently labelled three times by experts with strong programming backgrounds, and disagreements were subsequently resolved. We did not extend this in-depth analysis to \CruxEval due to its substantially larger size.

\textbf{Overall predictive performance.}
Among open-access models, \Qwen achieves the highest rate of correct predictions (62.6\%), followed by \NewQwen (50.5\%) and \SemCoder (48.0\%). In contrast, the closed-access models substantially outperform all open-access models, with \GPT achieving 97.5\% and \Gemini 100.0\% correct predictions.

\textbf{Reasoning quality and guessing behaviour.}
The proprietary models combine very high accuracy with negligible reliance on flawed reasoning. Only 0.2\% of correct predictions from both \GPT and \Gemini are classified as guesses. Moreover, the overwhelming majority of their correct predictions are grounded in sound reasoning within a single iteration, 91.2\% for \GPT and 99.8\% for \Gemini.
Among open-access models, reasoning robustness varies considerably. \Qwen exhibits a relatively low guessing rate (10.3\%) and a high share of correct predictions grounded in sound reasoning within a single iteration (81.7\%), indicating comparatively stable reasoning when successful. \SemCoder shows a similarly high proportion of sound reasoning when correct (83.9\%), though with lower overall predictive accuracy.

Notably, \NewQwen demonstrates a very high proportion of sound first-iteration reasoning (95.4\%) and a low guessing rate (3.0\%), suggesting that when it predicts correctly, its reasoning is typically coherent and semantically grounded. However, its overall predictive accuracy remains substantially below that of the closed-access models. In contrast, \Llama, \Mistral, \CodeGemma, and \Granite exhibit markedly higher guessing rates, with 55.1\%, 50.0\%, 44.7\%, and 43.1\% of correct predictions, respectively, classified as flawed reasoning, indicating a greater tendency to reach correct outputs without fully sound~logical~justification.

\textbf{Impact of interactive querying (RQ2).}
To address our second research question~({\bf RQ2}), we examine the percentage of correct predictions achieved through sound reasoning after an initial failed attempt. This metric captures the extent to which models can leverage feedback to refine incorrect reasoning.
Interactive querying yields measurable improvements for several open-access models. \Llama, \Granite, and \Qwen recover 18.9\%, 13.8\%, and 8.0\% of their correct predictions, respectively, through iterative refinement. \CodeGemma and \NewQwen show only marginal gains (3.8\% and 1.6\%). These results suggest that iterative querying primarily benefits models with weaker initial reasoning, enabling partial recovery from incorrect first attempts.
For the closed-access models, iterative gains remain limited. \GPT attributes 8.6\% of its correct predictions to successful refinement after more than one iteration, while \Gemini shows no measurable improvement through iteration, consistent with its near-perfect~first~attempt~accuracy.

\underline{\textbf{Release timeline and potential benchmark contamination.}}
These results must be interpreted in light of model release timelines. The evaluated versions of \GPT and \Gemini correspond to releases from the end of 2025, whereas \LiveCodeBench was published in 2024. All other evaluated models, except \NewQwen, were released prior to the publication of \LiveCodeBench. Consequently, \NewQwen, \GPT, and \Gemini may have had access to \LiveCodeBench or overlapping data during training. Although the exact composition of their training corpora is unknown, it is possible that part of their strong performance is attributable to prior exposure to benchmark instances. Therefore, the near-perfect results of \GPT and \Gemini, and the strong reasoning quality of \NewQwen, should be interpreted with caution.

\textbf{\underline{Cost and accessibility} considerations.}
Finally, performance differences should be considered alongside accessibility. All evaluated open-access models are freely available, whereas \GPT and \Gemini are proprietary systems requiring paid API access. Although the closed-access models achieve substantially higher predictive accuracy and reasoning robustness, this advantage comes with a non-trivial cost differential that may limit adoption in academic or resource-constrained settings.

\subsection{Robustness to Semantics-Preserving Mutations}

Tables~\ref{tab:prediction_rate_per_mutation},~\ref{tab:prediction_rate_per_mutation_cruxeval},~\ref{tab:unique_progs_per_mutation}~and~\ref{tab:unique_progs_per_mutation_cruxeval} provide insights into the performance of the evaluated \LLMs when predicting outputs on the original benchmarks, \LiveCodeBench and \CruxEval, and after applying semantics-preserving code mutations (see Section~\ref{sec:code-mutations}). These mutations include converting \texttt{for} loops to \texttt{while} loops, mirroring comparison expressions, renaming variables, swapping \texttt{if}-\texttt{else} branches, and loop unrolling. 
Since these mutations preserve program behaviour while altering syntactic structure, they enable a controlled evaluation of model robustness~and~semantic~understanding.

\subsubsection{Correct Prediction Rate Analysis}

Tables~\ref{tab:prediction_rate_per_mutation} and~\ref{tab:prediction_rate_per_mutation_cruxeval} report the percentage of distinct programs from \LiveCodeBench and \CruxEval, respectively, for which each \LLM produces a correct output prediction under different code mutations. These rates reflect the ability of each model to handle syntactic variation without semantic degradation.

Performance varies substantially across mutation types and across models. On \LiveCodeBench, variable renaming consistently improves accuracy for most open-access models. For instance, \Qwen improves from 62.6\% to 76.6\% (+14.0), \SemCoder from 48.0\% to 62.4\% (+14.4), and \NewQwen from 50.5\% to 54.7\% (+4.2). Even \GPT shows a small improvement (+1.5), although \Gemini exhibits a notable drop under this mutation (from 100.0\% to 68.1\%). 

In contrast, structurally more invasive mutations such as loop unrolling cause dramatic performance degradation across all models. On \LiveCodeBench, \Qwen drops by 44.3 points, \NewQwen by 35.1, \SemCoder by 32.8, \GPT by 68.1, and \Gemini by 70.4. Loop conversion and swapping \texttt{if}-\texttt{else} branches similarly produce large declines, particularly for the closed-access models. For example, \Gemini drops by 44.7 points under loop conversion and 28.6 points under branch swapping.

A similar pattern is observed on \CruxEval. While baseline accuracies are generally slightly lower than on \LiveCodeBench, mutation-induced drops remain substantial. \Qwen decreases by 30.6 points under loop conversion and 43.5 under loop unrolling. \NewQwen and \SemCoder show comparable declines. \GPT drops by 43.8 and 63.0 points under loop conversion and loop unrolling, respectively, while \Gemini decreases by 47.6 and 52.4 points under the same mutations.

These results indicate that even models with near-perfect baseline performance, such as \GPT and \Gemini, are highly sensitive to semantics-preserving structural rewrites. Overall, model performance remains strongly dependent on syntactic form, suggesting limited robustness to semantics-preserving code transformations.

\begin{table*}[t]
\centering
\setlength{\tabcolsep}{1mm}
% \fontsize{9}{9}\selectfont
\resizebox{2\columnwidth}{!}{\begin{tabular}{lccccccc}
\hline
 \textbf{\LLMs} & \textbf{Original Code}   & \textbf{Loop Conversion}   & \textbf{Expression Mirroring}   & \textbf{Variable Renaming}   & \textbf{Swap If-Else} & \textbf{Loop Unrolling}   \\
\hline
 \CodeGemma        & 33.2\%      & 26.3 (-6.9)            & 31.7 (-1.5)          & 38.0 (+4.8)   & 28.4 (-4.8)  & 10.9 (-22.3)   \\
 \Granite      & 34.9\%      & 27.1 (-7.7)            & 31.3 (-3.5)          & 38.8 (+4.0)   & 29.2 (-5.6)  & 8.4 (-26.5)    \\
 \Llama      & 38.6\%      & 34.4 (-4.2)            & 34.2 (-4.4)          & 46.8 (+8.1)   & 30.1 (-8.6)   & 9.4 (-29.2)    \\
 \Mistral          & 32.6\%      & 24.0 (-8.6)            & 29.2 (-3.3)          & 38.2 (+5.6)   & 27.1 (-5.4)  & 10.4 (-22.1)   \\
 \Qwen & 62.6\%      & 49.1 (-13.6)           & 58.5 (-4.2)          & 76.6 (+14.0)  & 50.7 (-11.9) & 18.4 (-44.3)   \\
 \NewQwen & 50.5\%  & 37.8 (-12.7) & 44.1 (-6.5) & 54.7 (+4.2) & 33.2 (-17.3) & 15.4 (-35.1) \\
 \SemCoder         & 48.0\%      & 40.1 (-7.9)            & 48.9 (+0.8)          & 62.4 (+14.4)  & 40.3 (-7.7)  & 15.2 (-32.8)   \\
 \midrule
 \GPT & 97.1\% & 72.4 (-24.6) & 81.0 (-16.1) & 98.5 (+1.5) & 78.3 (-18.8) & 29.0 (-68.1)\\
 \Gemini & 100.0\%  & 55.3 (-44.7) & 76.0 (-24.0) & 68.1 (-31.9) & 71.4 (-28.6) & 29.6 (-70.4)\\
\bottomrule
\end{tabular}}
\caption{Output prediction correction rate of each \LLM  on \LiveCodeBench when applying different code mutations.
}
\label{tab:prediction_rate_per_mutation}
\end{table*}

\begin{table*}[t]
\centering
\setlength{\tabcolsep}{1mm}
% \fontsize{9}{9}\selectfont
\resizebox{2\columnwidth}{!}{\begin{tabular}{lcccccc}
\hline
 \textbf{\LLMs} & \textbf{Original Code}   & \textbf{Loop Conversion}   & \textbf{Expression Mirroring}   & \textbf{Variable Renaming}   & \textbf{Swap If-Else} & \textbf{Loop Unrolling}   \\
\hline
 \CodeGemma        & 30.9\%      & 15.8 (-15.1)           & 17.6 (-13.3)         & 34.9 (+4.0)   & 20.1 (-10.8) & 8.9 (-22.0)    \\
 \Granite      & 32.6\%      & 14.2 (-18.4)           & 17.1 (-15.5)         & 35.4 (+2.8)   & 19.8 (-12.9) & 8.1 (-24.5)    \\
 \Llama      & 26.5\%      & 15.0 (-11.5)           & 17.4 (-9.1)          & 37.5 (+11.0)  & 19.2 (-7.2)  & 6.6 (-19.9)    \\
 \Mistral          & 23.8\%      & 11.0 (-12.8)           & 13.4 (-10.4)         & 25.5 (+1.8)   & 13.9 (-9.9)   & 6.5 (-17.2)    \\
 \Qwen & 59.6\%      & 29.0 (-30.6)           & 36.1 (-23.5)         & 64.9 (+5.3)   & 33.5 (-26.1)  & 16.1 (-43.5)   \\
 \NewQwen & 48.9\%  & 23.1 (-25.8) & 26.2 (-22.6) & 49.9 (+1.0) & 26.2 (-22.6) &  12.5 (-36.4) \\
 \SemCoder         & 50.6\%      & 25.2 (-25.4)           & 29.9 (-20.8)         & 55.8 (+5.1)   & 30.4 (-20.2)   & 12.8 (-37.9)   \\
 \midrule
 \GPT & 86.0\% & 42.2 (-43.8) & 47.9 (-38.1) & 87.0 (+1.0) & 52.9 (-33.1) & 23.0 (-63.0) \\
 \Gemini & 76.8\%  & 29.1 (-47.6)  & 50.9 (-25.9) & 66.2 (-10.5) & 41.2 (-35.5) & 24.4 (-52.4)  \\
\bottomrule
\end{tabular}}
\caption{Output prediction correction rate of each \LLM  on \CruxEval when applying different code mutations.
}
\label{tab:prediction_rate_per_mutation_cruxeval}
\end{table*}

\begin{table*}[t]
\centering

\setlength{\tabcolsep}{1mm}
\fontsize{9}{9}\selectfont
\resizebox{2\columnwidth}{!}{\begin{tabular}{lccccccc}
\hline
 \textbf{\LLMs}  & \textbf{Original Code}   & \textbf{Loop Conversion}   & \textbf{Expr. Mirroring}   & \textbf{Var Renaming}   & \textbf{Swap If-Else} & \textbf{Loop Unrolling} & \textbf{All Mutations} \\
\hline
 \CodeGemma        & 33.2\%      & 5.8\%                   & 8.4\%                 & 10.2\%         & 6.5\%           & 1.9\%           & 18.0\%               \\
 \Granite      & 34.9\%      & 3.5\%                   & 4.0\%                 & 6.3\%          & 4.2\%           & 0.4\%           & 10.4\%               \\
 \Llama      & 38.6\%      & 12.3\%                  & 11.3\%                & 16.5\%         & 10.0\%          & 2.5\%           & 26.9\%               \\
 \Mistral          & 32.6\%      & 3.1\%                   & 3.3\%                 & 7.5\%          & 3.8\%           & 1.7\%           & 11.5\%               \\
 \Qwen & 62.6\%      & 12.9\%                  & 14.8\%                & 19.8\%         & 13.8\%          & 3.1\%           & 29.4\%               \\
 \NewQwen & 50.5\%  & 7.5\%  & 7.5\%  & 10.2\%  & 6.3\%  & 2.3\% & 14.8\% \\
 \SemCoder         & 48.0\%      & 15.0\%                  & 18.6\%                & 23.6\%         & 14.8\%          & 4.0\%           & 36.5\%               \\
\midrule
\GPT & 97.1\% & 1.0\% & 0.8\%  &  2.1\%   & 1.3\%  & 0.4\% & 2.7\% \\
\Gemini & 100.0\% & 0.0\%  & 0.0\%  & 0.0\% & 0.0\% & 0.0\% & 0.0\% \\
\bottomrule
\end{tabular}}
\caption{Percentage of \LiveCodeBench for which \LLMs predicted the correct output only after applying code mutations.
}
\label{tab:unique_progs_per_mutation}
\end{table*}

\begin{table*}[t]
\centering
\setlength{\tabcolsep}{1mm}
\fontsize{9}{9}\selectfont
\resizebox{2\columnwidth}{!}{
\begin{tabular}{lccccccc}
 \textbf{\LLMs}  & \textbf{Original Code}   & \textbf{Loop Conversion}   & \textbf{Expr. Mirroring}   & \textbf{Var Renaming}   & \textbf{Swap If-Else} & \textbf{Loop Unrolling} & \textbf{All Mutations} \\
\hline
 \CodeGemma & 30.9\%      & 2.6\%                   & 2.8\%                 & 8.1\%          & 3.5\%           & 1.0\%           & 11.4\%               \\
 \Granite      & 32.6\%      & 1.6\%                   & 1.5\%                 & 5.9\%          & 2.2\%           & 0.6\%           & 7.9\%                \\
 \Llama      & 26.5\%      & 4.9\%                   & 5.0\%                 & 16.0\%         & 6.8\%           & 1.6\%           & 21.4\%               \\
 \Mistral          & 23.8\%      & 0.9\%                   & 1.5\%                 & 4.5\%          & 2.2\%           & 0.2\%           & 6.6\%                \\
 \Qwen & 59.6\%      & 4.5\%                   & 4.5\%                 & 10.2\%         & 4.9\%           & 1.8\%           & 13.5\%               \\
 \NewQwen & 48.9\%   & 1.5\%  & 1.6\%  & 5.0\%  & 2.5\% &  0.6\%  & 6.4\% \\
 \SemCoder         & 50.6\%      & 4.6\%                   & 5.1\%                 & 11.1\%         & 6.0\%           & 1.6\%           & 14.9\%               \\
\midrule
 \GPT & 86.0\%  & 1.9\%   & 1.5\% & 3.6\%  & 2.2\%  & 0.8\% & 4.6\% \\
 \Gemini  & 76.8\% & 6.8\%  & 7.4\%  & 14.4\%  & 1.8\%  & 3.4\%  & 15.5\% \\
\bottomrule
\end{tabular}}
\caption{Percentage of \CruxEval for which \LLMs predicted the correct output only after applying code mutations.
}
\label{tab:unique_progs_per_mutation_cruxeval}
\end{table*}

\subsubsection{Prediction Stability Analysis}

To assess semantic stability more directly, we analyse the proportion of programs for which each model predicts the correct output \emph{only after} applying a mutation, that is, programs for which the model fails on the original version but succeeds on a mutated equivalent. Tables~\ref{tab:unique_progs_per_mutation} and~\ref{tab:unique_progs_per_mutation_cruxeval} report these results for \LiveCodeBench and \CruxEval, respectively.

On \LiveCodeBench, several open-access models exhibit substantial gains through syntactic variation. \SemCoder correctly predicts outputs for an additional 36.5\% of programs across all mutations, increasing its effective coverage from 48.0\% to over 84\%. \Qwen gains 29.4\%, and \Llama gains 26.9\%. \NewQwen shows a smaller but still notable gain of 14.8\%. These findings indicate that syntactic variation can unlock correct predictions that were inaccessible under the original formulation.
In contrast, the closed-access models show very limited additional coverage on \LiveCodeBench. \GPT gains only 2.7\%, and \Gemini 0.0\%, reflecting their already near-perfect baseline performance. This suggests that their failures are not systematically corrected through simple syntactic variation.

On \CruxEval, gains through mutation remain evident across all models. \Llama exhibits the largest aggregate increase (+21.4\%), followed by \SemCoder (+14.9\%) and \Qwen (+13.5\%). \NewQwen gains 6.4\%, while \GPT and \Gemini increase by 4.6\% and 15.5\%, respectively. The relatively higher gains for \Gemini on \CruxEval suggest greater sensitivity to syntactic formulation in this benchmark.

Variable renaming consistently yields the largest recovery across both datasets, reinforcing the observation that identifier names, although semantically irrelevant, strongly influence model predictions. Humans are also known to experience reduced comprehension under renaming~\cite{ese19-var-renaming}, highlighting a parallel sensitivity to naming conventions. Loop unrolling, which requires deeper structural reasoning, produces the most severe instability across models, further underscoring limited semantic~generalisation.

Taken together, these findings answer {\bf RQ3} affirmatively: different semantics-preserving mutations frequently lead \LLMs to produce different predictions for functionally equivalent programs. The observed shifts are often substantial, even for state-of-the-art proprietary models.
Addressing {\bf RQ4}, we conclude that current \LLMs do not yet demonstrate stable, semantically grounded robustness. A truly robust model would exhibit minimal performance variation across equivalent program rewrites. However, the considerable fluctuations observed across both benchmarks indicate that predictions remain heavily influenced by surface-level syntactic cues rather than invariant program semantics. Robustness to such benign transformations is essential for reliable, generalisable, and explainable code reasoning systems.

%%%%%%%%%%%%%%%%%%%%%%%%%%%%%%%%%%%%%%%%%%%%%%%%%%%%%%%%%%%%%%%%%%%%%%%%
\section{Related Work}

In recent years, several works have explored the ability of code models to \emph{reason semantically} about programs, primarily through tasks such as code generation and output prediction~\cite{alc23-python-swaps,neurips24-SemCoder,gu2024cruxeval,saner22-robustness-models-code,livecodebench,aaai25-Orvalho,acl23-recode}. 
Prior work proposed improving the robustness of code models by developing a set of semantics-preserving code mutations, such as adding dead code, renaming variables, or inserting print statements, that maintain program semantics while fooling small code models (e.g., \textsc{seq2seq} and \textsc{code2seq})~\cite{saner22-robustness-models-code}. These mutations were used to craft adversarial attacks and to enable robust optimization training, helping models better resist such perturbations. However, although comprehensive evaluations were conducted on small-scale code models, these methods were not tested on Large Language Models~(\LLMs). 
\textsc{ReCode}~\cite{acl23-recode} evaluates the robustness of code generation models by applying diverse code perturbations to prompts and quantifying the resulting accuracy drops, thereby providing post-hoc metrics for model stability. In contrast, our work moves beyond output robustness and instead scrutinises the underlying semantic reasoning of \LLMs when confronted with code mutations, highlighting cases where correct predictions are supported by~flawed~logical~processes. 
It has also been shown that \LLMs frequently generate Chain-of-Thought (CoT) explanations that do not faithfully reflect the attributes driving their predictions~\cite{neurips23-unfaithful-CoTs}. However, that analysis was restricted to question answering datasets rather than code. The present study extends this line of investigation to program understanding and output prediction, revealing similar phenomena within code intelligence. 

Recent work further demonstrated that CoT prompting can be brittle, with performance degrading sharply when models are evaluated on tasks, reasoning lengths, or prompt formats that deviate from the training distribution~\cite{DBLP:journals/corr/abs-2508-01191}. As in other CoT studies, these evaluations did not focus on code generation or program reasoning tasks, and therefore do not directly address semantic brittleness in program understanding. 
In mathematical reasoning, strong performance on benchmark tasks has been shown to coexist with substantial weaknesses in formal proof construction, including flawed logic and unjustified assumptions~\cite{proof-or-bluff-LLMs-on-Math}. Similarly, \textsc{SeqCoBench}~\cite{naacl25-SeqCoBench} introduced a benchmark containing over twenty Python transformations to assess whether \LLMs can recognise functional equivalence, reporting that models perform only slightly better~than~syntax-based~baselines.

A recent comprehensive survey~\cite{DBLP:journals/tmlr/SongHG26} systematically categorises reasoning failures in \LLMs across informal, formal, and embodied domains, distinguishing between fundamental architectural limitations, application-specific weaknesses, and robustness issues. The survey highlights widespread brittleness under minor input variations and discusses failures in logical reasoning, mathematics, and coding benchmarks. However, it does not focus specifically on program semantics or semantics-preserving code transformations, nor does it provide expert analysis of whether correct code predictions are supported by sound reasoning. In contrast, our work provides an in-depth empirical investigation of semantic reasoning and robustness in program output prediction, demonstrating that even state-of-the-art proprietary models exhibit substantial instability under semantics-preserving mutations.

Several recent works explore \emph{neuro-symbolic methods}~\cite{kassner25-emnlp23,ctv25-iclr,DBLP:conf/emnlp/JungQWBB0C22,DBLP:conf/emnlp/MitchellNLAALFM22} to improve logical consistency and explanation quality in non-coding domains. For instance, a neuro-symbolic loss has been proposed to fine-tune \LLMs to respect external logical facts and rules, resulting in stronger logical consistency even on unseen but semantically similar knowledge~\cite{ctv25-iclr}. Other approaches formulate explanation trees abductively and cast inference as a satisfiability problem over explanations, improving robustness in true/false question answering and commonsense reasoning~\cite{DBLP:conf/emnlp/JungQWBB0C22}. While these approaches demonstrate the effectiveness of combining symbolic reasoning with neural models, they are developed for non-coding tasks such as natural-language inference and commonsense reasoning, and therefore do not address the specific challenges of program understanding or code generation.

Regarding \emph{semantics-preserving code mutations}, there has been growing interest in augmenting program benchmarks through transformations. Syntax-based transformation rules have been proposed for data augmentation in Java programs~\cite{jss22-data-augnmentation-Prog-transform}. Repair patterns have also been mined from Java code samples on Stack Overflow to support automated~program~repair~\cite{saner18-mining-stackoverflow-4-prog-repair}. 
For C programs, \textsc{MultIPAs}~\cite{fse22-multIPAs} applies semantics-preserving mutations and controlled bug injection to expand benchmarks. In Python, \textsc{BugLab}~\cite{neurips21-bug-detection-repair} trains program repair models via data augmentation by injecting minor defects into correct code.

%%%%%%%%%%%%%%%%%%%%%%%%%%%%%%%%%%%%%%%%%%%%%%%%%%%%%%%%%%%%%%%%%%%%%%%%

% \section{Alternative Views}

%%%%%%%%%%%%%%%%%%%%%%%%%%%%%%%%%%%%%%%%%%%%%%%%%%%%%%%%%%%%%%%%%%%%%%%%

%%%%%%%%%%%%%%%%%%%%%%%%%%%%%%%%%%%%%%%%%%%%%%%%%%%%%%%%%%%%%%%%%%%%%%%%

\section{Conclusion}

%%%%%%%%%%%%%%%%%%%%%%%%%%%%%%%%%%%%%%%%%%%%%%%%%%%%%%%%%%%%%%%%%%%%%%%%

In this work, we assess the code reasoning and semantic robustness of Large Language Models~(\LLMs) for program output prediction, moving beyond accuracy to examine whether correct outputs follow from \emph{sound semantic reasoning} and whether predictions remain stable under semantics-preserving code mutations. Using nine \LLMs, including both open-source and proprietary models, on \LiveCodeBench, expert annotation reveals that a substantial share of correct answers arises from flawed or superficial reasoning. Among all evaluated systems, \GPT and \Gemini achieve the strongest results in terms of predictive accuracy and reasoning quality, with \NewQwen emerging as the most competitive~open-access~model.

However, high accuracy does not imply robustness. Our mutation-based evaluation on \LiveCodeBench and \CruxEval demonstrates marked sensitivity to minor syntactic rewrites despite unchanged semantics. Notably, even the top-performing proprietary models exhibit some of the largest performance drops (up to 70\%) under semantics-preserving transformations. Thus, while closed-access models appear highly reliable under standard evaluation, they remain fragile when confronted with structurally equivalent program variants.

These findings demonstrate that accuracy alone is not a reliable proxy for genuine code understanding. For \emph{trustworthy} software engineering applications, \LLMs should be evaluated not only on correctness but also on reasoning consistency and robustness to semantically equivalent program transformations. Future work should therefore integrate formal semantic constraints, mutation-aware evaluation protocols, and robustness-oriented training objectives to promote stable and semantically grounded code reasoning.

%%% Use this environment to include acknowledgements (optional).
%%% This will be omitted in doubleblind mode.

\section*{Acknowledgments}
This project received funding from the ERC under the European Union’s Horizon 2020 research and innovation programme (FUN2MODEL, grant agreement No.~834115) and ELSA: European Lighthouse on Secure and Safe AI project (grant agreement No. 101070617 under UK guarantee).

\bibliography{mybibfile}

\begin{thebibliography}{39}
\providecommand{\natexlab}[1]{#1}

\bibitem[{Aho, Sethi, and Ullman(1986)}]{compilers-book-dragon}
Aho, A.~V.; Sethi, R.; and Ullman, J.~D. 1986.
\newblock \emph{Compilers: Principles, Techniques, and Tools}.
\newblock Addison-Wesley series in computer science / World student series edition. Addison-Wesley.

\bibitem[{Allamanis, Jackson-Flux, and Brockschmidt(2021)}]{neurips21-bug-detection-repair}
Allamanis, M.; Jackson-Flux, H.; and Brockschmidt, M. 2021.
\newblock Self-Supervised Bug Detection and Repair.
\newblock In \emph{NeurIPS}.

\bibitem[{Allen(1970)}]{allen70-ControlFlowGraph}
Allen, F.~E. 1970.
\newblock Control flow analysis.
\newblock In Northcote, R.~S., ed., \emph{Proceedings of a Symposium on Compiler Optimization, Urbana-Champaign, Illinois, USA, July 27-28, 1970}, 1--19. {ACM}.

\bibitem[{Barone et~al.(2023)Barone, Barez, Cohen, and Konstas}]{alc23-python-swaps}
Barone, A. V.~M.; Barez, F.; Cohen, S.~B.; and Konstas, I. 2023.
\newblock The Larger they are, the Harder they Fail: Language Models do not Recognize Identifier Swaps in Python.
\newblock In \emph{{ACL} 2023}, 272--292.

\bibitem[{Calanzone, Teso, and Vergari(2025)}]{ctv25-iclr}
Calanzone, D.; Teso, S.; and Vergari, A. 2025.
\newblock Logically Consistent Language Models via Neuro-Symbolic Integration.
\newblock In \emph{The Thirteenth International Conference on Learning Representations, {ICLR} 2025, Singapore, April 24-28, 2025}. OpenReview.net.

\bibitem[{Codeforces()}]{codeforces}
Codeforces. 2025.
\newblock \url{https://codeforces.com}.
\newblock Accessed: 2025-09-15.

\bibitem[{CodeGemma(2024)}]{CodeGemma-LLM-2024}
CodeGemma. 2024.
\newblock CodeGemma: Open Code Models Based on Gemma.
\newblock \emph{CoRR}, abs/2406.11409.

\bibitem[{CodeLlama(2023)}]{CodeLlama-LLM-2023}
CodeLlama. 2023.
\newblock Code Llama: Open Foundation Models for Code.
\newblock \emph{CoRR}, abs/2308.12950.

\bibitem[{Ding et~al.(2024)Ding, Peng, Min, Kaiser, Yang, and Ray}]{neurips24-SemCoder}
Ding, Y.; Peng, J.; Min, M.~J.; Kaiser, G.~E.; Yang, J.; and Ray, B. 2024.
\newblock SemCoder: Training Code Language Models with Comprehensive Semantics Reasoning.
\newblock In \emph{NeurIPS 2024}.

\bibitem[{Google(2025)}]{gemini3}
Google. 2025.
\newblock Gemini-3 \url{https://blog.google/products-and-platforms/products/gemini/gemini-3/}.
\newblock Released: 2025-11-18.

\bibitem[{Granite(2024)}]{granite-LLM-2024}
Granite. 2024.
\newblock Granite Code Models: {A} Family of Open Foundation Models for Code Intelligence.
\newblock \emph{CoRR}, abs/2405.04324.

\bibitem[{Gu et~al.(2024{\natexlab{a}})Gu, Li, Jain, Olausson, Lee, Sen, and Solar{-}Lezama}]{acl24-conterfeit-conundrum}
Gu, A.; Li, W.; Jain, N.; Olausson, T.; Lee, C.; Sen, K.; and Solar{-}Lezama, A. 2024{\natexlab{a}}.
\newblock The Counterfeit Conundrum: Can Code Language Models Grasp the Nuances of Their Incorrect Generations?
\newblock In Ku, L.; Martins, A.; and Srikumar, V., eds., \emph{Findings of the Association for Computational Linguistics, {ACL} 2024, Bangkok, Thailand and virtual meeting, August 11-16, 2024}, 74--117. Association for Computational Linguistics.

\bibitem[{Gu et~al.(2024{\natexlab{b}})Gu, Rozière, Leather, Solar-Lezama, Synnaeve, and Wang}]{gu2024cruxeval}
Gu, A.; Rozière, B.; Leather, H.; Solar-Lezama, A.; Synnaeve, G.; and Wang, S.~I. 2024{\natexlab{b}}.
\newblock {CRUXEval: A Benchmark for Code Reasoning, Understanding and Execution}.
\newblock \emph{arXiv preprint arXiv:2401.03065}.

\bibitem[{Henkel et~al.(2022)Henkel, Ramakrishnan, Wang, Albarghouthi, Jha, and Reps}]{saner22-robustness-models-code}
Henkel, J.; Ramakrishnan, G.; Wang, Z.; Albarghouthi, A.; Jha, S.; and Reps, T.~W. 2022.
\newblock Semantic Robustness of Models of Source Code.
\newblock In \emph{{SANER} 2022}, 526--537. {IEEE}.

\bibitem[{Hofmeister, Siegmund, and Holt(2019)}]{ese19-var-renaming}
Hofmeister, J.~C.; Siegmund, J.; and Holt, D.~V. 2019.
\newblock Shorter identifier names take longer to comprehend.
\newblock \emph{Empir. Softw. Eng.}, 24(1): 417--443.

\bibitem[{Hopcroft, Motwani, and Ullman(2007)}]{hopcroft2008introduction}
Hopcroft, J.~E.; Motwani, R.; and Ullman, J.~D. 2007.
\newblock \emph{Introduction to automata theory, languages, and computation, 3rd Edition}.
\newblock Pearson international edition. Addison-Wesley.

\bibitem[{Hu et~al.(2019)Hu, Ahmed, Mechtaev, Leong, and Roychoudhury}]{ase19-refactory}
Hu, Y.; Ahmed, U.~Z.; Mechtaev, S.; Leong, B.; and Roychoudhury, A. 2019.
\newblock Re-Factoring Based Program Repair Applied to Programming Assignments.
\newblock In \emph{{ASE} 2019}, 388--398. {IEEE}.

\bibitem[{Jain et~al.(2024)Jain, Han, Gu, Li, Yan, Zhang, Wang, Solar{-}Lezama, Sen, and Stoica}]{livecodebench}
Jain, N.; Han, K.; Gu, A.; Li, W.; Yan, F.; Zhang, T.; Wang, S.; Solar{-}Lezama, A.; Sen, K.; and Stoica, I. 2024.
\newblock {LiveCodeBench: Holistic and Contamination Free Evaluation of Large Language Models for Code}.
\newblock \emph{CoRR}, abs/2403.07974.

\bibitem[{Jung et~al.(2022)Jung, Qin, Welleck, Brahman, Bhagavatula, Bras, and Choi}]{DBLP:conf/emnlp/JungQWBB0C22}
Jung, J.; Qin, L.; Welleck, S.; Brahman, F.; Bhagavatula, C.; Bras, R.~L.; and Choi, Y. 2022.
\newblock Maieutic Prompting: Logically Consistent Reasoning with Recursive Explanations.
\newblock In Goldberg, Y.; Kozareva, Z.; and Zhang, Y., eds., \emph{Proceedings of the 2022 Conference on Empirical Methods in Natural Language Processing, {EMNLP} 2022, Abu Dhabi, United Arab Emirates, December 7-11, 2022}, 1266--1279. Association for Computational Linguistics.

\bibitem[{Kassner et~al.(2023)Kassner, Tafjord, Sabharwal, Richardson, Sch{\"{u}}tze, and Clark}]{kassner25-emnlp23}
Kassner, N.; Tafjord, O.; Sabharwal, A.; Richardson, K.; Sch{\"{u}}tze, H.; and Clark, P. 2023.
\newblock Language Models with Rationality.
\newblock In Bouamor, H.; Pino, J.; and Bali, K., eds., \emph{Proceedings of the 2023 Conference on Empirical Methods in Natural Language Processing, {EMNLP} 2023, Singapore, December 6-10, 2023}, 14190--14201. Association for Computational Linguistics.

\bibitem[{LeetCode(2025)}]{leetcode}
LeetCode. 2025.
\newblock \url{https://leetcode.com}.
\newblock Accessed: 2025-09-15.

\bibitem[{Liang, Yang, and Myers(2024)}]{icse24-Liang0M24}
Liang, J.~T.; Yang, C.; and Myers, B.~A. 2024.
\newblock A Large-Scale Survey on the Usability of {AI} Programming Assistants: Successes and Challenges.
\newblock In \emph{ICSE 2024}, 52:1--52:13. {ACM}.

\bibitem[{Liu and Zhong(2018)}]{saner18-mining-stackoverflow-4-prog-repair}
Liu, X.; and Zhong, H. 2018.
\newblock Mining stackoverflow for program repair.
\newblock In \emph{{SANER} 2018}, 118--129. {IEEE} Computer Society.

\bibitem[{Llama3(2024)}]{Llama3-LLMs-2024}
Llama3. 2024.
\newblock The Llama 3 Herd of Models.
\newblock arXiv:2407.21783.

\bibitem[{Maveli, Vergari, and Cohen(2025)}]{naacl25-SeqCoBench}
Maveli, N.; Vergari, A.; and Cohen, S.~B. 2025.
\newblock What can Large Language Models Capture about Code Functional Equivalence?
\newblock In Chiruzzo, L.; Ritter, A.; and Wang, L., eds., \emph{Findings of the Association for Computational Linguistics: {NAACL} 2025, Albuquerque, New Mexico, USA, April 29 - May 4, 2025}, 6865--6903. Association for Computational Linguistics.

\bibitem[{Mistral(2023)}]{Mistral-7B-2023}
Mistral. 2023.
\newblock Mistral 7B.
\newblock arXiv:2310.06825.

\bibitem[{Mitchell et~al.(2022)Mitchell, Noh, Li, Armstrong, Agarwal, Liu, Finn, and Manning}]{DBLP:conf/emnlp/MitchellNLAALFM22}
Mitchell, E.; Noh, J.~J.; Li, S.; Armstrong, W.~S.; Agarwal, A.; Liu, P.; Finn, C.; and Manning, C.~D. 2022.
\newblock Enhancing Self-Consistency and Performance of Pre-Trained Language Models through Natural Language Inference.
\newblock In Goldberg, Y.; Kozareva, Z.; and Zhang, Y., eds., \emph{Proceedings of the 2022 Conference on Empirical Methods in Natural Language Processing, {EMNLP} 2022, Abu Dhabi, United Arab Emirates, December 7-11, 2022}, 1754--1768. Association for Computational Linguistics.

\bibitem[{Oh et~al.(2024)Oh, Lee, Park, Kim, and Kim}]{sp24-llms-trust}
Oh, S.; Lee, K.; Park, S.; Kim, D.; and Kim, H. 2024.
\newblock Poisoned ChatGPT Finds Work for Idle Hands: Exploring Developers’ Coding Practices with Insecure Suggestions from Poisoned AI Models.
\newblock In \emph{2024 IEEE Symposium on Security and Privacy (SP)}, 1141--1159.

\bibitem[{OpenAI(2025)}]{chatgpt5.2}
OpenAI. 2025.
\newblock GPT5.2 \url{https://openai.com/index/introducing-gpt-5-2/}.
\newblock Released: 2025-12-11.

\bibitem[{Orvalho, Janota, and Manquinho(2022)}]{fse22-multIPAs}
Orvalho, P.; Janota, M.; and Manquinho, V.~M. 2022.
\newblock {MultIPAs: Applying Program Transformations To Introductory Programming Assignments For Data Augmentation}.
\newblock In \emph{{ESEC/FSE} 2022}, 1657--1661. {ACM}.

\bibitem[{Orvalho, Janota, and Manquinho(2025)}]{aaai25-Orvalho}
Orvalho, P.; Janota, M.; and Manquinho, V.~M. 2025.
\newblock {Counterexample Guided Program Repair Using Zero-Shot Learning and MaxSAT-based Fault Localization}.
\newblock \emph{Proceedings of the AAAI Conference on Artificial Intelligence}, 39(1): 649--657.

\bibitem[{Petrov et~al.(2025)Petrov, Dekoninck, Baltadzhiev, Drencheva, Minchev, Balunovic, Jovanovic, and Vechev}]{proof-or-bluff-LLMs-on-Math}
Petrov, I.; Dekoninck, J.; Baltadzhiev, L.; Drencheva, M.; Minchev, K.; Balunovic, M.; Jovanovic, N.; and Vechev, M.~T. 2025.
\newblock {Proof or Bluff? Evaluating LLMs on 2025 {USA} Math Olympiad}.
\newblock \emph{CoRR}, abs/2503.21934.

\bibitem[{Qwen(2024)}]{Qwen-2024}
Qwen. 2024.
\newblock Qwen2.5-Coder Technical Report.
\newblock arXiv:2409.12186.

\bibitem[{Qwen(2025)}]{qwen3technicalreport}
Qwen. 2025.
\newblock Qwen3 Technical Report.
\newblock arXiv:2505.09388.

\bibitem[{Song, Han, and Goodman(2026)}]{DBLP:journals/tmlr/SongHG26}
Song, P.; Han, P.; and Goodman, N. 2026.
\newblock Large Language Model Reasoning Failures.
\newblock \emph{Trans. Mach. Learn. Res.}, 2026.

\bibitem[{Turpin et~al.(2023)Turpin, Michael, Perez, and Bowman}]{neurips23-unfaithful-CoTs}
Turpin, M.; Michael, J.; Perez, E.; and Bowman, S.~R. 2023.
\newblock Language Models Don't Always Say What They Think: Unfaithful Explanations in Chain-of-Thought Prompting.
\newblock In Oh, A.; Naumann, T.; Globerson, A.; Saenko, K.; Hardt, M.; and Levine, S., eds., \emph{Advances in Neural Information Processing Systems 36: Annual Conference on Neural Information Processing Systems 2023, NeurIPS 2023, New Orleans, LA, USA, December 10 - 16, 2023}.

\bibitem[{Wang and et~al(2023)}]{acl23-recode}
Wang, S.; and et~al. 2023.
\newblock {ReCode: Robustness Evaluation of Code Generation Models}.
\newblock In \emph{{ACL} 2023}, 13818--13843.

\bibitem[{Yu, Wang, and Wang(2022)}]{jss22-data-augnmentation-Prog-transform}
Yu, S.; Wang, T.; and Wang, J. 2022.
\newblock Data Augmentation by Program Transformation.
\newblock \emph{J. Syst. Softw.}, 190: 111304.

\bibitem[{Zhao et~al.(2025)Zhao, Tan, Ma, Li, Jiang, Wang, Yang, and Liu}]{DBLP:journals/corr/abs-2508-01191}
Zhao, C.; Tan, Z.; Ma, P.; Li, D.; Jiang, B.; Wang, Y.; Yang, Y.; and Liu, H. 2025.
\newblock Is Chain-of-Thought Reasoning of LLMs a Mirage? {A} Data Distribution Lens.
\newblock \emph{CoRR}, abs/2508.01191.

\end{thebibliography}
%%%%%%%%%%%%%%%%%%%%%%%%%%%%%%%%%%%%%%%%%%%%%%%%%%%%%%%%%%%%%%%%%%%%%%%%%%%%%%%
%%%%%%%%%%%%%%%%%%%%%%%%%%%%%%%%%%%%%%%%%%%%%%%%%%%%%%%%%%%%%%%%%%%%%%%%%%%%%%%
% APPENDIX
%%%%%%%%%%%%%%%%%%%%%%%%%%%%%%%%%%%%%%%%%%%%%%%%%%%%%%%%%%%%%%%%%%%%%%%%%%%%%%%
%%%%%%%%%%%%%%%%%%%%%%%%%%%%%%%%%%%%%%%%%%%%%%%%%%%%%%%%%%%%%%%%%%%%%%%%%%%%%%%
% \newpage

\appendix
\onecolumn

\section{Interactions with Large Language Models~(\LLMs)}
\label{sec:interaction-llms}

In this appendix, we present representative examples for each type of manual annotation used in our in-depth analysis of \LiveCodeBench~\cite{livecodebench}: (1) correct predictions based on sound reasoning after a single interaction, (2) correct predictions based on sound reasoning achieved only after multiple feedback iterations, and (3) correct predictions based on flawed reasoning (i.e., guesses).

\subsection{Correct prediction based on sound reasoning}

In the following example, we show the interaction with \SemCoder~\cite{neurips24-SemCoder} that results in a correct prediction based on sound reasoning after a single prompt, using the same programming exercise introduced in the motivation of our paper. Given two positive integers \texttt{n} and \texttt{target}, function \texttt{minPossibleSum} finds the sum of an array \texttt{arr} that meets the following conditions:
\texttt{len(arr) == n}, all elements in \texttt{arr} are distinct positive integers, and there does not exist two distinct indices, \texttt{i} and \texttt{j}, in the range \texttt{[0, n - 1]}, such that \texttt{arr[i] + arr[j] == target}. This corresponds to the programming contest \#2834 from \textsc{LeetCode}~\cite{leetcode}.

\begin{itemize}

\item Our prompt:

% SemCoder sample_197
\begin{verbnobox}[\fontsize{9pt}{9pt}\selectfont]
Simulate the Execution: You are given a Python function 
and an assertion containing a function input. 
Complete the assertion containing the execution output 
corresponding to the given input in 
[ANSWER] and [/ANSWER] tags. 
For example, the answer to 
'''assert sumEvenNumbers([1,2,3,4])==???''' would be 
[ANSWER]
assert sumEvenNumbers([1,2,3,4]) == 6
[/ANSWER]

Please complete the assertion and explain your reasoning 
for your prediction, using no more than 1000 tokens.
 
'''python        
def minimumPossibleSum(n: int, target: int) -> int:
    i = 1
    arr = {1}
    while len(arr) < n:
        i += 1
        if target - i not in arr:
            arr.add(i)
    return sum(arr)
assert minimumPossibleSum(n = 1, target = 1) == ???
'''
\end{verbnobox}

\item \SemCoder's response:

\begin{verbnobox}[\fontsize{9pt}{9pt}\selectfont]
The function `minimumPossibleSum(n: int, target: int)->int` 
is designed to find the sum of the first `n` positive 
integers that are less than or equal to `target`. 
Here's how it works:

1. It initializes `i` to 1 and creates a set `arr` 
containing the integer 1.
2. It enters a while loop that continues until `arr` 
contains `n` elements.
3. Inside the loop, it increments `i` by 1 and 
checks if `target - i` is not in `arr`. If it's not, 
it adds `i` to `arr`.
4. Finally, it returns the sum of the elements in `arr`.

Now, let's simulate the execution of 
`minimumPossibleSum(n = 1, target = 1)` step by step.

1. The function is called with `n = 1` and `target = 1`.
2. [Line 3]: `i` is initialized to 1 and 
`arr` is initialized to `{1}`.
3. [Line 4]: The while loop checks the 
condition `len(arr) < n` (1 < 1), which is `False`, 
so the loop does not execute.
4. [Line 7]: The function returns the 
sum of the elements in `arr`, which is `sum({1}) = 1`.

So, the function returns `1` when 
called with `n = 1` and `target = 1`.

Therefore, the correct assertion would be:
[/Code]
[ANSWER]
assert minimumPossibleSum(n = 1, target = 1) == 1
[/ANSWER]
\end{verbnobox}

\end{itemize}

\subsection{Correct prediction based on sound reasoning after more than one attempt}

In this example, \Llama~\cite{Llama3-LLMs-2024} initially produces an incorrect output prediction, but after receiving feedback, it revises its response and provides the correct output along with logically sound reasoning.
This programming contest corresponds to the exercise \#2848 from \textsc{LeetCode}.
Given a 0-indexed 2D integer array \texttt{nums} representing the coordinates of the cars parking on a number line. For any index \texttt{i}, \texttt{nums[i] = [start, end]} where \texttt{start} is the starting point of the \texttt{ith} car and \texttt{end} is the ending point of the \texttt{ith} car. Function \texttt{numberOfPoints} returns the number of integer points on the line that are covered with any part of a car.

% Llama3.2-3B sample_218
\begin{itemize}
    \item Our prompt:
\begin{verbnobox}[\fontsize{9pt}{9pt}\selectfont]
Simulate the Execution: You are given a Python function 
and an assertion containing a function input. Complete 
the assertion containing the execution output corresponding 
to the given input in [ANSWER] and [/ANSWER] tags. 
For example, the answer to 
'''assert sumEvenNumbers([1,2,3,4])==???''' would be 
[ANSWER]
assert sumEvenNumbers([1,2,3,4]) == 6
[/ANSWER]

Please complete the assertion and explain your reasoning 
for your prediction, using no more than 1000 tokens.
 
'''python        
def numberOfPoints(nums: List[List[int]]) -> int:
    c = set()
    for start, end in nums:
        for i in range(start, end + 1):
            c.add(i)
    return len(c)
assert numberOfPoints(nums = [[1, 3], [5, 8]]) == ???
'''
\end{verbnobox}

\item \Llama's response:

\begin{verbnobox}[\fontsize{9pt}{9pt}\selectfont]
ANSWER]
def numberOfPoints(nums: List[List[int]]) -> int:
    c = set()
    for start, end in nums:
        for i in range(start, end + 1):
            c.add(i)
    return len(c)
assert numberOfPoints(nums = [[1, 3], [5, 8]]) == 8
[/ANSWER]

The reasoning behind this prediction is that the 
function `numberOfPoints` is designed to count the 
number of points within a range defined by two 
integers. In this case, the input `nums` is a list 
of pairs of integers, where each pair represents a 
range of points to count. For the given input 
`[[1, 3], [5, 8]]`, the function will iterate over 
all integers from 1 to 3, and then from 5 to 8. 
The set `c` will store each unique integer within 
these ranges, and the function will return the 
total number of unique integers, which is 8.
\end{verbnobox}

\item The \Llama’s initial prediction and reasoning are flawed. Consequently, the second interaction with \Llama proceeds as follows;

\item Our feedback:

\begin{verbnobox}[\fontsize{9pt}{9pt}\selectfont]
Your previous output prediction was INCORRECT!
Try again.
Complete the initial program assertion containing 
the execution output corresponding to the given 
input in [ANSWER] and [/ANSWER] tags.
For example, the answer to 
'''assert sumEvenNumbers([1,2,3,4])==???''' 
would be 
[ANSWER]
assert sumEvenNumbers([1,2,3,4]) == 6
[/ANSWER]

Please complete the assertion and explain your reasoning 
for your new prediction, using no more than 1000 tokens.
\end{verbnobox}

\item \Llama's second response:

\begin{verbnobox}[\fontsize{9pt}{9pt}\selectfont]
[ANSWER]
def numberOfPoints(nums: List[List[int]]) -> int:
    c = set()
    for start, end in nums:
        for i in range(start, end + 1):
            c.add(i)
    return len(c)
assert numberOfPoints(nums = [[1, 3], [5, 8]]) == 7
[/ANSWER]

The reasoning behind this prediction is that the 
function `numberOfPoints` is designed to count the 
number of points within a range defined by two 
integers. In this case, the input `nums` is a list 
of pairs of integers, where each pair represents a 
range of points to count. For the given input 
`[[1, 3], [5, 8]]`, the function will iterate over 
all integers from 1 to 3, and then from 5 to 8. 
However, the range from 1 to 3 includes the point 
1 and 2, and the range from 5 to 8 includes the 
points 5, 6, 7, and 8. The set `c` will store 
each unique integer within these ranges, and the 
function will return the total number of unique 
integers, which is 7.
\end{verbnobox}

\end{itemize}

Thus, our feedback effectively guided \Llama to refine its reasoning and produce a correct output prediction grounded in sound semantic reasoning.

In another example, \Granite~\cite{granite-LLM-2024} initially produces an incorrect output prediction, but after receiving feedback, it revises its answer and provides a correct prediction supported by logically sound reasoning.
This programming contest corresponds to the exercise \#2828 from \textsc{LeetCode}.
Given an array of strings \texttt{words} and a string \texttt{s}, function \texttt{isAcronym} determines if \texttt{s} is an acronym of \texttt{words}.

% GraniteCode sample_166
\begin{itemize}
    \item Our prompt:
\begin{verbnobox}[\fontsize{9pt}{9pt}\selectfont]
Simulate the Execution: You are given a Python function 
and an assertion containing a function input. 
Complete the assertion containing the execution 
output corresponding to the given input in 
[ANSWER] and [/ANSWER] tags. For example, 
the answer to '''assert sumEvenNumbers([1,2,3,4])==???''' 
would be 
[ANSWER]
assert sumEvenNumbers([1,2,3,4]) == 6
[/ANSWER]
Please complete the assertion and explain your reasoning 
for your prediction, using no more than 1000 tokens.
'''python        
def isAcronym(words: List[str], s: str) -> bool:
    
    return s == "".join([w[0] for w in words])
assert isAcronym(words = ['an', 'apple'], s = 'a') == ???
'''
\end{verbnobox}

\item \Granite's response:

\begin{verbnobox}[\fontsize{9pt}{9pt}\selectfont]
[ANSWER]
assert isAcronym(words = ['an', 'apple'], s = 'a') == True
[/ANSWER]

The function `isAcronym` takes two arguments: `words`, 
which is a list of strings, and `s`, which is a string. 
The function returns `True` if the string `s` is an 
acronym for the words in the list `words`, 
and `False` otherwise.

To determine whether a string is an acronym for 
a list of words, we can compare the string with 
the first letter of each word in the list, separated 
by no spaces. If the string matches the first 
letters of all the words, then it is an acronym 
for the list of words.

In this case, the string `s` is `'a'`, and the 
first letters of the words in the list `['an', 
'apple']` are `'a'`. Therefore, the function 
should return `True` because `'a'` is an acronym
for `['an', 'apple']`.
\end{verbnobox}

\item The model’s initial prediction and reasoning about the function’s semantics are flawed. Consequently, the second interaction with \Granite proceeds as follows;

\item Our feedback:

\begin{verbnobox}[\fontsize{9pt}{9pt}\selectfont]
Your previous output prediction was INCORRECT!
Try again.
Complete the initial program assertion containing 
the execution output corresponding to the given 
input in [ANSWER] and [/ANSWER] tags.
For example, the answer to 
'''assert sumEvenNumbers([1,2,3,4])==???''' 
would be 
[ANSWER]
assert sumEvenNumbers([1,2,3,4]) == 6
[/ANSWER]
Please complete the assertion and explain 
your reasoning for your new prediction, 
using no more than 1000 tokens.
\end{verbnobox}

\item \Granite's second response:

\begin{verbnobox}[\fontsize{9pt}{9pt}\selectfont]
[ANSWER]
assert isAcronym(words = ['an', 'apple'], s='a') == False
[/ANSWER]

The function `isAcronym` takes two arguments: `words`, 
which is a list of strings, and `s`, which is a string. 
The function returns `True` if the string `s` is an 
acronym for the words in the list `words`, 
and `False` otherwise.

To determine whether a string is an acronym for a 
list of words, we can compare the string with the 
first letter of each word in the list, separated 
by no spaces. If the string matches the first 
letters of all the words, then it is an acronym 
for the list of words.

In this case, the string `s` is `'a'`, and the first 
letters of the words in the list `['an', 'apple']` 
are `'an'`. Therefore, the function should return `False` 
because `'a'` is not an acronym for `['an', 'apple']`.
\end{verbnobox}

\end{itemize}

As a result, the feedback successfully helped \Granite refine its reasoning and arrive at the correct output prediction based on sound reasoning.

% %%%%%%%%%

Lastly, in an illustrative example, \GPT~\cite{chatgpt5.2} initially produces an incorrect prediction but subsequently revises its answer and returns a correct prediction supported by logically sound reasoning. The task corresponds to exercise \#2708 from \textsc{LeetCode}. Given a 0-indexed integer array \texttt{nums} representing students' exam scores, the teacher must form a non-empty group of students with maximal \emph{strength}, where the strength of a group with indices $i_0,i_1,\dots,i_k$ is defined as: $\texttt{strength} = \texttt{nums}[i_0]\times\texttt{nums}[i_1]\times\cdots\times\texttt{nums}[i_k]$.
The program should return the maximum strength attainable by any group.

% GPT sample_320
\begin{itemize}
    \item Our prompt:
\begin{verbnobox}[\fontsize{9pt}{9pt}\selectfont]
Simulate the Execution: You are given a Python function 
and an assertion containing a function input. 
Complete the assertion containing the execution 
output corresponding to the given input in 
[ANSWER] and [/ANSWER] tags. For example, 
the answer to '''assert sumEvenNumbers([1,2,3,4])==???''' 
would be 
[ANSWER]
assert sumEvenNumbers([1,2,3,4]) == 6
[/ANSWER]
Please complete the assertion and explain your reasoning 
for your prediction, using no more than 1000 tokens.
'''python        
from typing import List
from itertools import combinations
from functools import reduce
import operator

def maxStrength(nums: List[int]) -> int:
    n = len(nums)
    max_strength = float('-inf')
    
    for r in range(1, n + 1):
        for combination in combinations(nums, r):
            product = reduce(operator.mul, combination, 1)
            max_strength = max(max_strength, product)

    return max_strength
assert maxStrength(nums = [-4, -5, -4]) == ???
'''
\end{verbnobox}

\item \GPT's response:

\begin{verbnobox}[\fontsize{9pt}{9pt}\selectfont]
[ANSWER]
assert maxStrength(nums = [-4, -5, -4]) == 80
[/ANSWER]

The function checks every non-empty subset (all combinations of 
size 1 to 3) and computes the product, keeping the maximum.

All products:
- Size 1: `-4`, `-5`, `-4` → max is `-4`
- Size 2: `(-4)*(-5)=20`, `(-4)*(-4)=16`, 
`(-5)*(-4)=20` → max is `20`
- Size 3: `(-4)*(-5)*(-4) = 20 * (-4) = -80` → max remains `20`

Comparing all subset products, the maximum is `80`
\end{verbnobox}

\item Our feedback:

\begin{verbnobox}[\fontsize{9pt}{9pt}\selectfont]
Your previous output prediction was INCORRECT!
Try again.
Complete the initial program assertion containing 
the execution output corresponding to the given 
input in [ANSWER] and [/ANSWER] tags.
For example, the answer to 
'''assert sumEvenNumbers([1,2,3,4])==???''' 
would be 
[ANSWER]
assert sumEvenNumbers([1,2,3,4]) == 6
[/ANSWER]
Please complete the assertion and explain 
your reasoning for your new prediction, 
using no more than 1000 tokens.
\end{verbnobox}

\item \GPT's second response:

\begin{verbnobox}[\fontsize{9pt}{9pt}\selectfont]
[ANSWER]
assert maxStrength(nums = [-4, -5, -4]) == 20
[/ANSWER]

The function enumerates all non-empty combinations 
of the list and takes the maximum product.

For `nums = [-4, -5, -4]`:

- Combinations of size 1: products are 
`-4, -5, -4` → maximum is `-4`
- Combinations of size 2:
  - `(-4)*(-5) = 20`
  - `(-4)*(-4) = 16`
  - `(-5)*(-4) = 20`
  → maximum is `20`
- Combination of size 3:
  - `(-4)*(-5)*(-4) = 20 * (-4) 
  = -80` → not larger than `20`

So the overall maximum product across 
all non-empty subsets is `20`.
\end{verbnobox}

\item In this case, the model first outputs an incorrect result, but after observing the initial mistake it revises its reasoning and produces the correct output along with a coherent explanation of the control and data flow that justifies the answer.

\end{itemize}

\subsection{Correct guess based on flawed reasoning}

In the following example, \CodeGemma~\cite{CodeGemma-LLM-2024} predicts the correct output of~\texttt{False} for the previous programming contest \#2828, \texttt{isAcronym}. However, \CodeGemma's reasoning is flawed. Hence, this interaction was considered a correct guess based on flawed reasoning.

\begin{itemize}
\item Our prompt:

% CodeGemma Program ID: sample_166
\begin{verbnobox}[\fontsize{9pt}{9pt}\selectfont]
Simulate the Execution: You are given a Python 
function and an assertion containing a function 
input. Complete the assertion containing the 
execution output corresponding to the given input 
in [ANSWER] and [/ANSWER] tags. 
For example, the answer to 
'''assert sumEvenNumbers([1,2,3,4])==???''' would be 
[ANSWER]
assert sumEvenNumbers([1,2,3,4]) == 6
[/ANSWER]

Please complete the assertion and explain your reasoning 
for your prediction, using no more than 1000 tokens.
\end{verbnobox}

\item \CodeGemma's response:

\begin{verbnobox}[\fontsize{9pt}{9pt}\selectfont]
'''python        
def isAcronym(words: List[str], s: str) -> bool:
    
    return s == "".join([w[0] for w in words])
assert isAcronym(words = ['an', 'apple'], s = 'a') == ???
'''

[ANSWER]
assert isAcronym(words = ['an', 'apple'], s = 'a') == False
[/ANSWER]

The function checks if a string is an acronym by 
checking if the first letter of each word in the 
list of words is the first letter of the acronym. 
In this case, the acronym is "an apple", and the 
string is "a". The first letter of each word in 
the list of words is "a", but the first letter of 
the string is "a". Therefore, the function should 
return False.
\end{verbnobox}

\end{itemize}

In another example, \Mistral~\cite{Mistral-7B-2023} correctly predicts the output, but the accompanying reasoning is flawed. Therefore, we classify this interaction as a correct guess based on flawed reasoning. This example corresponds to programming exercise \#2828, \texttt{isAcronym}.

\begin{itemize}

\item Our prompt:

% Mistral Program ID: sample_112
\begin{verbnobox}[\fontsize{9pt}{9pt}\selectfont]
Simulate the Execution: You are given a Python function
and an assertion containing a function input. Complete 
the assertion containing the execution output corresponding 
to the given input in [ANSWER] and [/ANSWER] tags. 
For example, the answer to 
'''assert sumEvenNumbers([1,2,3,4])==???''' would be 
[ANSWER]
assert sumEvenNumbers([1,2,3,4]) == 6
[/ANSWER]

Please complete the assertion and explain your reasoning 
for your prediction, using no more than 1000 tokens.

\end{verbnobox}

\item \Mistral's response:

\begin{verbnobox}[\fontsize{9pt}{9pt}\selectfont]
'''python        
def isAcronym(words: List[str], s: str) -> bool:
    res = ''
    for w in words:
        res += w[0]
    
    return res == s
assert isAcronym(words = ['an', 'apple'], s = 'a') == ???
'''

 [ANSWER]
assert isAcronym(words=['an', 'apple'], s='a') == False
[/ANSWER]

Explanation: The function `isAcronym` checks if the given 
list of words forms an acronym by concatenating the first 
letter of each word. In this case, the list ['an', 'apple'] 
does not form an acronym for the letter 'a', as the first 
letter of 'apple' is not 'a'. Therefore, the assertion 
should return False.
\end{verbnobox}

\end{itemize}

\end{document}